\documentclass[aps,prd,twocolumn,nofootinbib,superscriptaddress]{revtex4-1}
\usepackage[toc,page]{appendix}

\usepackage{graphicx}
\usepackage{psfrag}
\usepackage{color}
\usepackage{amsmath}
\usepackage{amsfonts}
\usepackage{amssymb}
\usepackage{textcomp}
\usepackage{multirow}
\usepackage[breaklinks=true,colorlinks=true,linkcolor=blue,urlcolor=blue,citecolor=blue]{hyperref}

\usepackage[normalem]{ulem}

\newcommand{\mcomment}[1]{}

\graphicspath{{Plots/}}

\begin{document}
	
	
	\title {Thermodynamically consistent treatment of repulsive corrections in HRG}
	\author{Somenath Pal}
	\affiliation{Variable Energy Cyclotron Centre, 1/AF, Bidhan Nagar , Kolkata-700064, India}
		\email{somenathpalphysics@gmail.com}
	
	
	\def\be{\begin{equation}}
		\def\ee{\end{equation}}
	\def\bearr{\begin{eqnarray}}
		\def\eearr{\end{eqnarray}}

	\begin{abstract}
    We reformulate the treatment of density-dependent chemical potential shifts appearing in excluded-volume implementations of the hadron resonance gas model. An auxiliary classical representation is constructed in which a common energy shift is determined by preserving the scalar number density, ensuring thermodynamic consistency. Hadron radii are parametrized through a liquid-drop inspired mass–radius relation with two parameters: the pion radius and a scaling exponent. The resulting framework reproduces lattice QCD results for lower-order conserved-charge susceptibilities at zero chemical potentials with only two adjustable parameters.
	\end{abstract}
	\maketitle
	
	\section{Introduction}
	Strongly interacting matter in thermodynamic equilibrium is a hot topic of research for some decades. It is directly linked to our endeavour for a better understanding of the constituent particles found in the Universe and the interplay of the fundamental forces acting among them. Such matter is expected to have existed microseconds after the Big Bang~\cite{kolb2018early} and may be found in the core of the neutron stars~\cite{rajagopal2001frontier}. It can also be produced in relativistic heavy-ion collision experiments. To describe a collection of interacting particles, one should have the knowledge of the strength of the interactions acting between each particle pair in the system. However, the nonperturbative nature of strong interaction in the low momentum transfer regime makes it challenging to theoretically describe the properties of a collection of strongly interacting particles. The random motion of the particles in the system adds to the complication as it requires the knowledge of the position, momentum and the forces on all the particles simultaneously at a particular time point. However, the problem simplifies significantly for a collection of particles in thermodynamic equilibrium enabling the investigation of the collective system without any microscopic details. Hence, the study of the in-equilibrium thermodynamic phases and phase diagram of strong interactions at high temperatures and densities is of significant importance for proper understanding of the strongly interacting systems both found in nature and produced experimentally.
    Unfortunately, in spite of rigorous efforts in the theoretical and experimental explorations, the phase diagram of strongly interacting matter is far from complete. Lattice QCD (LQCD) is used as a benchmark to describe the QCD matter at vanishing chemical potentials within the limit of certain discretization error, especially due to lower mass of pions in the hadronic sector~\cite{PhysRevD.95.054504,PhysRevD.95.094503,PhysRevD.105.074511}.  Lattice QCD (LQCD) suffers from the sign problem for non-zero chemical potentials and are not expected to yield reliable results if the value of chemical potential is significant as compared to the temperature.
    Effective model approach provides a simple yet effective way of producing important insights and approximate quantitative results of strongly interacting matter under certain energy regimes. Important aspects of such matter has been explored with the extremely successful Hadron Resonance Gas (HRG) model~\cite{Dashen:1969ep,pisarski1982strings,KARSCH200367,EJIRI2006275,HUOVINEN201026,PhysRevD.86.034509,PhysRevC.109.055206,Kahangirwe:2024xyl}, which incorporates attractive interactions among hadrons by considering the unstable resonances as stable particles. Repulsive interactions have been incorporated in the HRG model in the form of excluded volume correction~\cite{Rischke:1991ke,Steinheimer:2010ib}, van der Waals interaction approach~\cite{Vovchenko:2020lju}, mean-field repulsive interaction~\cite{Kapusta:1982qd,Pal:2023zzp} providing partial success to describe the lattice data at zero chemical potentials.
	
	The limitations of the mentioned variants of HRG model may be attributed to the following reasons: (1) The S-matrix formulation of statistical mechanics developed by Dashen, Ma and Bernstein~\cite{Dashen:1969ep} provides a formally exact expression for the thermodynamic potential in terms of scattering phase shift. The approach captures attractive interactions via resonances. However, its practical realization in the HRG model relies on truncation of the hadronic spectrum and neglect of repulsive contributions. Phenomenological implementations of repulsion via excluded-volume corrections introduce effective shifts of the chemical potentials, which are not derived from the S-matrix framework. These shifts introduce model-dependent contributions to higher-order derivatives of the pressure. Consequently, fluctuation observables become particularly sensitive to the chosen implementation of repulsion. In other words, with excluded-volume implementations, the chemical potential is modified in a density-dependent manner. This leads to higher-order derivatives of pressure (susceptibilities) receiving contributions from the phenomenological density dependence of the effective chemical potential. 
    This motivates a reformulation of density-dependent chemical potential shifts.   
    
	(2) the thermodynamic properties are highly sensitive to the size of the hadrons. Unfortunately, excluded volume of each and every hadron species is not known yet. Although there is a consensus regarding the value of the pion radius to be of the order of 0.2 $fm$, no conclusive results exist for most of the other hadrons.
	
	An attempt is therefore made in this paper to circumvent the concern no. (1) We introduce an auxiliary classical representation in which a common energy shift is determined by requiring equality of the scalar number density. The repulsive interactions have been incorporated by the excluded volume approach. Excluded volumes of the hadrons have been calculated from the pion size through a simple liquid drop modelling of the hadrons.
\section{Framework}
The effect of interactions is to shift the individual particle energy levels, the magnitude of the shift for the species $i$ is denoted by $\nu_i(T,\mu_B,\mu_Q,\mu_S,\{\mathcal{G}_i\})$, where, $\nu_i$ depends on intrinsic properties $\{\mathcal{G}_i\}$ of hadrons,
i.e., hadron size, shape, spin, etc. In this work, we consider the shift in the energy levels due to the excluded volume only. 
\subsection{Excluded volume correction}
The actual number density becomes
	
	\begin{eqnarray}
	n &=& \Sigma_i n_i  = \Sigma_i \frac{g_i}{2\pi^2} \nonumber \\
    &\times &\int_0^{\infty} \frac{p^2 dp}{e^{(\sqrt{p^2 + m_i^2} - \mu_i + \nu_i(T,\mu_B,\mu_Q,\mu_S,\{\mathcal{G}_i\}))/T} \pm 1}
		\label{total_number_density_1}
	\end{eqnarray}

    To calculate the energy level $\mathcal{E}$ in the classical statistics picture, we need to know $\nu_i(T,\mu_B,\mu_Q,\mu_S,\{\mathcal{G}_i\})$ for every hadron species $i$ at $(T, \mu_B, \mu_Q, \mu_S)$. For simplicity, we assume the hadrons to be spherical. Using the excluded volume model~\cite{Rischke:1991ke,Begun:2012rf}, $\nu_i(T,\mu_B,\mu_Q,\mu_S,\{\mathcal{G}_i\})=P v_i$, where, $v_i=\frac{16}{3}\pi( r_i+r'(T,\mu_B,\mu_Q,\mu_S))^3$ is the excluded volume of hadronic species $i$ assuming a spherical shape of the hadrons. Till date, very little progress has been achieved regarding the calculation and measurement of hadron radii. This limitation hinders the direct use of the physical size of the hadrons as an intrinsic property in any model calculation. However, some works suggest that the pion radius is somewhere close to 0.2 fm with a large uncertainty in the actual numerical value~\cite{Krutov:2024adh,Feng:2019geu}. In this work, we take the pion radius $r_{\pi}$ = 0.2 fm and use the liquid drop model~\cite{Polyakov:2002yz} to calculate the radii of other hadrons in terms of the pion radius. The contribution of the spin of the hadrons has been considered through the spin degeneracy $g_i$ only. Thus, the radius introduced here should be interpreted as an effective mechanical radius characterizing the spatial distribution of internal forces. It is not the electromagnetic charge radius. Motivated by liquid-drop considerations of confined systems, one expects the radius to scale with mass as a power law. The average radius $r'$ is obtained by first calculating the total physical volume occupied by all hadrons ($\Sigma_i n_i \frac{4}{3}\pi r_i^3$), and dividing by the total number density to obtain the mean volume per particle. The corresponding effective radius is defined as that of a sphere with this average volume. This provides a mean-field approximation for a multicomponent hard-sphere mixture: $r'(T,\mu_B,\mu_Q,\mu_S)=\Big (\frac{\sum_i n_i r_i^3}{\sum_i n_i} \Big )^{1/3}$. We take $r_i=\sqrt{\frac{3}{5}}R_h$ as a phenomenological effective mechanical radius inspired by liquid-drop considerations.
    \subsection{Conceptual framework to facilitate the calculation}
	  The shift $\nu_i$ in the energy levels of individual hadron species $i$, appearing in equation~(\ref{total_number_density_1}) needs to be evaluated. 
The shift $\nu_i$ in the energy levels of individual hadron species $i$, appearing in equation~(\ref{total_number_density_1}) needs to be evaluated. The treatment of the unstable resonances as stable particles in the HRG model is just an effective way to mimic the long range attractive interactions and should lead to improper results when the effective chemical potential is nonzero. The particle content of the system remaining unchanged with time, we can imagine that all the hadrons undergo elastic collisions among themselves with the conservation of energy and momenta in each collision.
So, we construct an auxiliary classical representation to facilitate the calculation. This is a computational reparametrization, not a physical replacement of quantum statistics. We determine a single shift $\mathcal{E}$ such that the scalar number density is preserved, achieving a classical mapping corresponding to the actual system. Higher thermodynamic derivatives inherit corrections through the implicit $\mu$-dependence of $\mathcal{E}$. Thus, we have simplified the problem from calculation of $N$ unknowns, $\nu_i$ in the quantum statistics picture to the calculation of a single unknown $\mathcal{E}$ only in the equivalent classical statistics picture. To find the relationship between the expressions of the same thermodynamic quantities in these two pictures, we first note that the scalar number density in the actual quantum statistics is equal to that in the auxiliary picture with MB statistics. Hence, we can write

	  \begin{equation}
	  	n_{actual} = \Sigma_i n_{i,actual}  = n_{auxiliary} = \Sigma_i n_{i,auxiliary}
	  \end{equation}
	  
	  As per our argument, the number density in the auxiliary frame can be written as
	  \begin{equation}
	  	n = \Sigma_i n_i = \Sigma_i \frac{g_i}{2\pi^2} m_i^2 T \mathcal{K}_2 \Big (\frac{m_i}{T}\Big ) e^{\frac{\mu_i+\mathcal{E}}{T}}
	  	\label{total_number_density_2}
	  \end{equation}
	  Comparing equations~(\ref{total_number_density_1}) and ~(\ref{total_number_density_2}), we can calculate $\mathcal{E}$ numerically.
	  
	  Here, $g_i$ is the spin degeneracy of hadronic species $i$; $m_i$ is the mass, $\mathcal{K}_2$ is the modified Bessel functions. The Pressure $(P)$  of system is evaluated in the original quantum formulation.
	  \begin{align}
P &= \sum_i p_i= \sum_i \left( \pm \frac{g_i}{2\pi^2} \right) \nonumber \\
     &\times \int_0^{\infty} p^2 \, dp \,
     \log \Bigg[
     1 \pm 
     \exp\!\left(
     -\frac{\sqrt{p^2+m_i^2}
     - \mu_i + \nu_i}{T}
     \right)
     \Bigg].
\label{pressure}
\end{align}
	  $\mathcal{E}$ is determined implicitly from number density equality in the two pictures using the equations (\ref{total_number_density_1}), (\ref{total_number_density_2}) and (\ref{pressure}). The relation $n=\frac{\partial P}{\partial \mu}$ is numerically satisfied. All susceptibilities are computed using total derivatives including the implicit $\mu$-dependence of $\mathcal{E}$. The mapping to an auxiliary Maxwell–Boltzmann representation is motivated by kinetic theory. In a multicomponent gas, the collision integral preserves the scalar number density in the local rest frame. A uniform shift of single-particle energies corresponds to a common additive term in the exponent of the distribution functions. In the classical limit, this shift factors out of the momentum integral and modifies the overall normalization of the distribution without altering its momentum dependence. Therefore, matching the scalar number density provides a natural and minimal condition for determining the common energy shift $\mathcal{E}$. Other thermodynamic quantities such as pressure or energy density depend on additional momentum weights and would introduce redundant constraints. The scalar density is thus the appropriate quantity for establishing equivalence between the quantum-statistical formulation with density-dependent shifts and the auxiliary classical representation.

    \subsection{Hadron radius parametrization}
A spherical liquid drop is a simple structure which can be used to model the size of the hadrons. We consider a constant inner pressure $P_0(R)$ at each and every point $r$ inside a hadron with radius $R$ except at the surface. Hence, the pressure and the shear force distribution in the liquid drop becomes~\cite{Polyakov:2002yz}
	  
	  \begin{eqnarray}
	  	P(r)&=&P_0(R)\Theta (R-r) -\frac{P_0(R) R}{3}\delta (r-R)\ ; \nonumber \\
        S(r)&=&\gamma \delta (r-R)
	  	\label{liquid_drop_equation}
	  \end{eqnarray}
	   Here, $S(r)$ is the surface tension, $\gamma$ is the surface tension coefficient and Kelvin relation gives $P_0(R)=\frac{2 \gamma}{R}$~\cite{Thomson}.
	   
	   The stress tensor of a liquid drop is
	   \begin{equation}
	   	T^{ij}(r)=\Big (\frac{r^i r^j}{r^2}-\frac{1}{3}\delta^{ij} \Big ) S(r) + \delta^{ij}P(r)
	   \end{equation}
	   For a spherically symmetric drop, the above equation gives the normal force $F_r$ at a distance r from the centre of the drop (The tangential force is zero for perfectly spherical drop):
	   \begin{equation}
	   	\frac{dF_r}{dS_r}=\frac{2}{3}S(r)+P(r)
	   \end{equation}
   Which gives the force $F_r$ at the surface of a sphere of radius $r$ of mass $m$
   \begin{equation}
   	F_r= \int \Big(\frac{2}{3}S(r)+P(r)\Big) 8 \pi r dr = 4\pi P_0 (r) r^2
   \end{equation} In the last step, we have used eq.~(\ref{liquid_drop_equation}). We now consider the liquid drop being prepared in the vacuum from zero initial mass by gradually expanding the liquid drop to the radius $R$ of a hadron species of mass $m$. We phenomenologically identify the integrated pressure work with the hadron mass, as a liquid-drop inspired ansatz, with $R$ as the radius and the mass $m$ of the hadron:
    \begin{equation}
    	\int_0^R 4\pi P_0(x) x^2 dx = m = \frac{4}{3}\pi \rho(R) R^3=\frac{4}{3}\pi k^{\prime}(R) P_0(R) R^3
    	\label{radius_equation}
    \end{equation}
   where mass density
   \begin{equation}
   \rho(R) = k^{\prime}(R) P_0(R)
   \label{mass_density}
   \end{equation}
   Differentiating eq.~(\ref{radius_equation}) we have,
   \begin{equation}
   P_0(R) R^2 = k(R) \frac{\partial P_0(R)}{\partial R}
   + P_0(R) \frac{\partial k(R)}{\partial R}
    \label{diff_radius_equation}
   \end{equation}
where $k(R) = \frac{1}{3} \pi R^3 k^{\prime}(R)$. The above differential equation is subject to the boundary conditions: $k(R) P_0(R)=0$ for $R=\infty$ and $R=0$, since no hadron having these values of radius exists. Although $P_0(R)$ should be a well-behaved function defining a physical quantity, the same may not be true for $k(R)$, which was introduced to express the density as a function of $P_0(R)$, although the exact functional form is unknown. A functional dependence of the form~\ref{mass_density} is expected to produce reliable results provided $k(R)$ is properly chosen. The condition $k(R) P_0(R)=0$ for $R=\infty$ implies that only the form $k(R)=k_0 R^{-3}$ is consistent with the equation~(\ref{diff_radius_equation}), resulting into
\begin{equation}
	\frac{\partial P_0(R)}{\partial R} + \frac{3-k_0}{3k_0} \frac{P_0(R)}{R}=0
	\label{diff_2}
\end{equation}
   The above equation has solution of the form $P_0(R)=C R^{\frac{k_0-3}{3k_0}}$. Imposing the condition $P_0(R_{pi})=C R_{pi}^{\frac{k_0-3}{3k_0}}$ we have
   \begin{equation}
   	R_h = R_{\pi} \Big ( \frac{m_{\pi}}{m_h} \Big )^{\frac{3k_0}{3+8k_0}} = R_{\pi} \Big ( \frac{m_{\pi}}{m_h} \Big )^{\mathcal{A}}
   \end{equation}
The mechanical radius of a hadron is given by $r_i=\sqrt{\frac{3}{5}}R_h$, inspired from the liquid drop model~\cite{polyakov}.

	\vspace{4mm} 
	
\section{Results}
\begin{figure}[t]
       \includegraphics[width=\columnwidth]{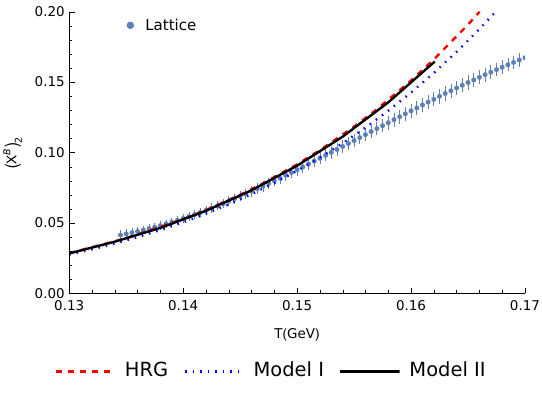}\\
       \includegraphics[width=\columnwidth]{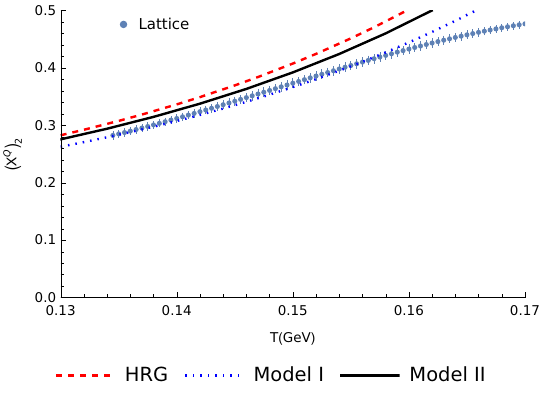}\\
       \includegraphics[width=\columnwidth]{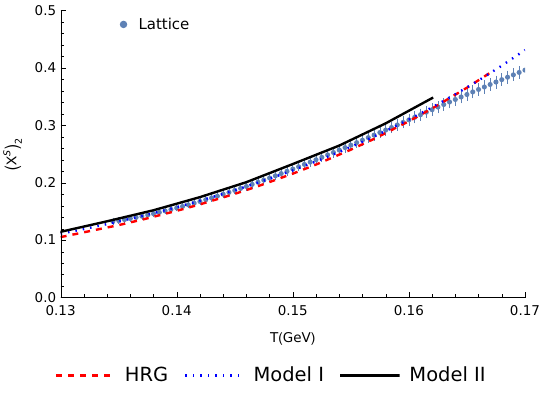}
       \caption{Second order baryon number ($\chi^B_2$), electric charge ($\chi^Q_2$) and strangeness ($\chi^S_2$) susceptibilities at zero chemical potentials. Lattice data has been taken from Ref.~\cite{Bollweg:2021vqf}.}
       \label{fig_second}
\end{figure}

\begin{figure}[t]
       \includegraphics[width=\columnwidth]{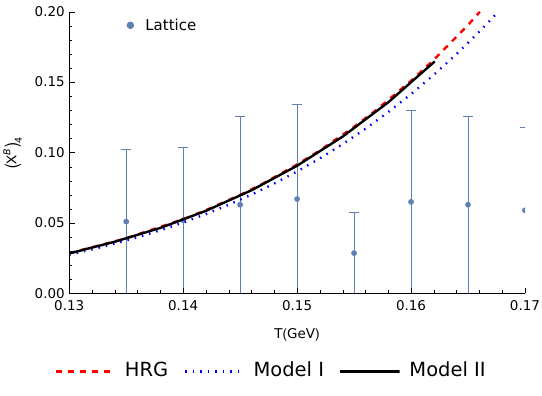}\\
       \includegraphics[width=\columnwidth]{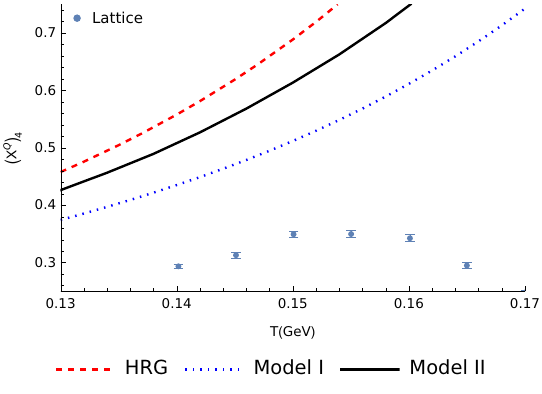}\\
       \includegraphics[width=\columnwidth]{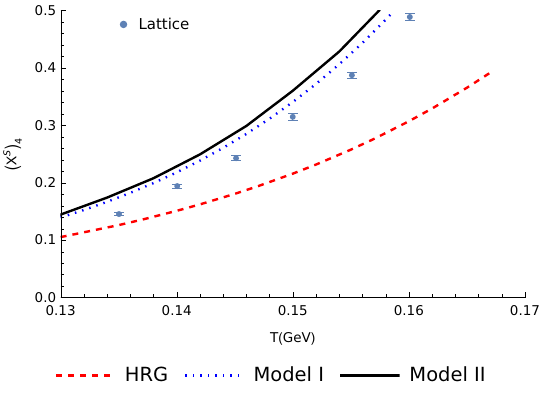}
       \caption{Fourth order baryon number ($\chi^B_4$), electric charge ($\chi^Q_4$) and strangeness ($\chi^S_4$) susceptibilities at zero chemical potentials. Lattice data has been taken from Ref.~\cite{Borsanyi:2018grb}.}
       \label{fig_fourth}
\end{figure}
\begin{figure}[t]
\includegraphics[width=\columnwidth]{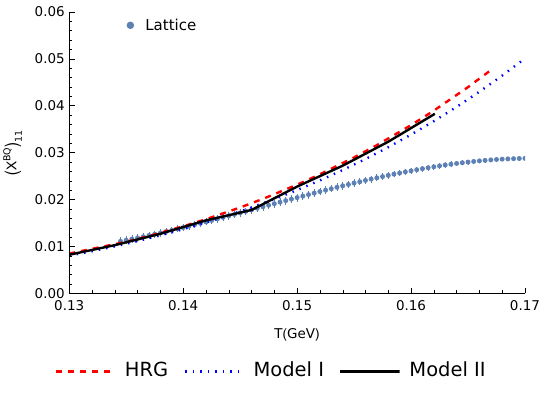}\\
\includegraphics[width=\columnwidth]{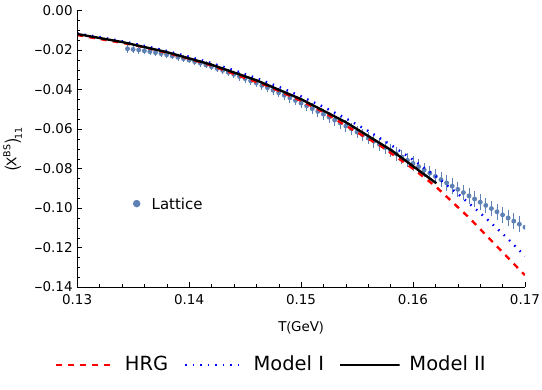}\\
\includegraphics[width=\columnwidth]{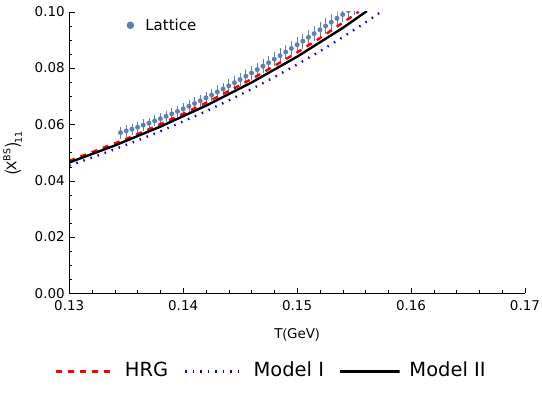}
\caption{Crossed susceptibilities of conserved charges at zero chemical potentials. Lattice data has been taken from Ref.~\cite{Karthein:2021cmb}.}
\label{fig_crossed}
\end{figure}
\begin{figure}[t]
      \includegraphics[width=\columnwidth]{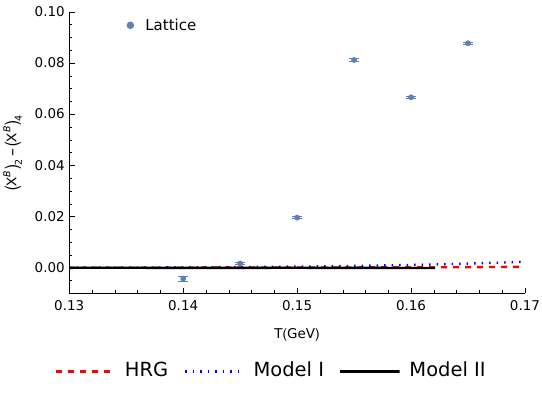}\\
       \includegraphics[width=\columnwidth]{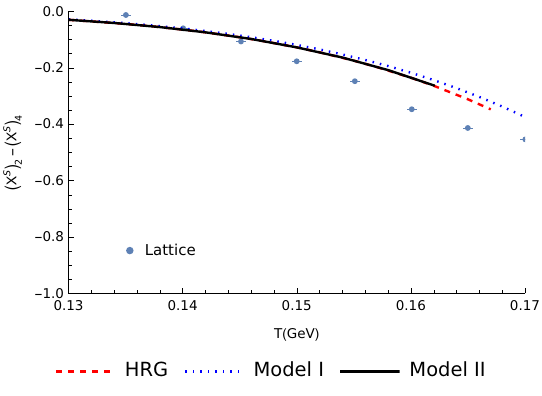}
\caption{Difference of second and fourth order baryon number ($\chi^B_2$) and strangeness ($\chi^S_2$) susceptibilities at zero chemical potentials. Lattice data has been taken from Ref.~\cite{Bellwied:2015lba,Borsanyi:2018grb}.}
\label{fig_diff}
\end{figure}
\begin{figure}[t]
       \includegraphics[width=\columnwidth]{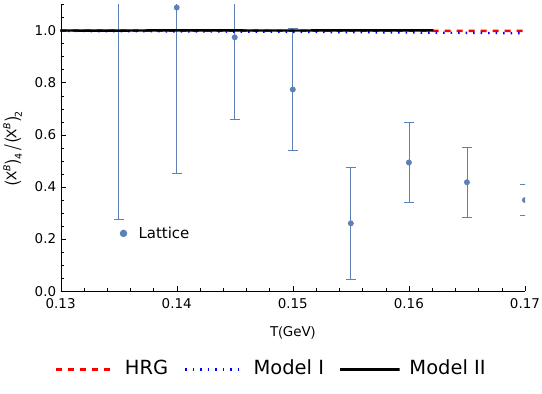}\\
       \includegraphics[width=\columnwidth]{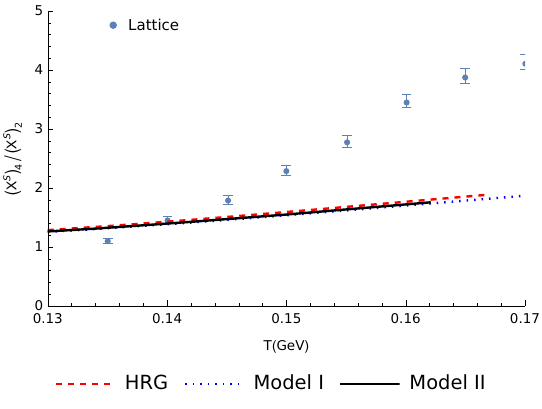}
       \caption{Ratios of second and fourth order baryon number ($\chi^B_2$) and strangeness ($\chi^S_2$) susceptibilities at zero chemical potentials. Lattice data has been taken from Ref.~\cite{Borsanyi:2023wno,Borsanyi:2018grb}.}
       \label{fig_ratio}
\end{figure}
Susceptibilities of conserved charges are defined as
\begin{equation}
\chi^{BQS}_{lmn}(T,\mu_B,\mu_Q,\mu_S)
=
\frac{\partial^{\,l+m+n}}
{\partial\left(\frac{\mu_B}{T}\right)^l
 \partial\left(\frac{\mu_Q}{T}\right)^m
 \partial\left(\frac{\mu_S}{T}\right)^n}
\left(\frac{P}{T^4}\right).
\label{susceptibility_definition}
\end{equation}
To show the importance of the proper treatment of the non-zero chemical potential term in the effective free particle picture, we compare the conserved charge susceptibilities under two different scenario: (1) Model I with the chemical potential term, arising due to repulsive interactions, absorbed in the individual particle energy level, as described in the previous section of this paper. (2) Model II with the traditional treatment of the effective chemical potential, where no equivalent effective particle picture has been considered. 
We show the results for pion radius $r_{\pi}$ = 0.2 $fm$ and $\mathcal{A}$=3 and investigate the conserved charge susceptibilities at zero chemical potentials. In this work, we
use the QMHRG2020 list of hadrons~\cite{bollweg2021second}, which includes all the
confirmed (3-and 4-star baryon resonances) and the unconfirmed (1- and 2-star baryon resonances as
well as mesons not listed in the PDG 2020 summary tables) hadrons. Additionally, this list also includes quark-model states in the strange and non-strange baryon sectors. All the baryon number, electric charge and strangeness susceptibilities, except the fourth order electric charge susceptibility, are well reproduced within the "Model I" except $\chi_Q^4$. It is worth mentioning that the lattice calculations of the fourth-order electric charge susceptibility  exhibits comparatively larger lattice uncertainties. Agreement in the higher order strange sector is also not good which may be due to the undiscovered high mass strange hadrons. The charge susceptibilities of crossed order containing baryon number as one of the conserved charge also show good agreement with the lattice data.
\section{Summary}
In this work, we reformulate the treatment of density-dependent chemical potential shifts appearing in excluded-volume implementations by mapping the quantum statistics picture to an equivalent classical statistics picture and assigning a common ground state energy level for all the hadron species. Thermodynamic consistency in the classical statistics picture is maintained by equating the number densities in both pictures. Assuming the hadrons as classical liquid drops, a phenomenological formula for the hadron radius varying as power law over mass has been derived. We find that the resulting model satisfactorily reproduces the lattice data for all the lower order conserved charge susceptibilities simultaneously with only two model parameters: pion radius $R_{\pi}=0.2\ fm$ and $\mathcal{A}=3$.

It is expected that this simple model is not able to describe all the higher susceptibilities because the calculation of excluded volumes of the hadronic species based on the spherical liquid drop model is an oversimplified treatment. Discovery of more higher mass particles in the future will help us improve the results.

	 \section{Acknowledgment}
	SP acknowledges financial support from the Department of Atomic Energy, India.

	





	\bibliographystyle{apsrev4-1}
	\bibliography{paper}
	
\end{document}